\documentclass[epj]{svjour}
\usepackage[loading]{tracefnt}

\usepackage[intlimits]{amsmath}
\usepackage{amssymb}
\usepackage{longtable}
\usepackage{multirow}
\usepackage{setspace}
\usepackage{graphicx}
\usepackage{textcomp}         
\usepackage{array}
\usepackage{nicefrac}
\usepackage{slashed}
\usepackage[usenames,table]{xcolor}
\usepackage[colorlinks=true,linkcolor=blue,citecolor=blue,urlcolor=blue]{hyperref}
\usepackage[numbers,sort&compress]{natbib}

\newcommand{\gS}[1]{#1\!\!\!\!\!\not~}

\newcommand{\pslash}{\gS{p}~}

\newcommand{\ONE}{ {\bf 1} }

\begin{document}

\title{Mass spectra and Regge trajectories of light mesons in the Bethe-Salpeter approach}

\author{Christian S. Fischer\thanks{\email{christian.fischer@theo.physik.uni-giessen.de}}
 \and Stanislav Kubrak\thanks{\email{stanislav.kubrak@theo.physik.uni-giessen.de}} \and Richard Williams\thanks{\email{richard.williams@theo.physik.uni-giessen.de}}}
\institute{Institut f\"ur Theoretische Physik, Justus-Liebig--Universit\"at Giessen, 35392 Giessen, Germany.}

\date{Received: date / Revised version: date}

\abstract{
We extend the calculation of relativistic bound-states of a fermion anti-fermion pair in the
Bethe-Salpeter formalism to the case of total angular momentum $J=3$. Together with results for $J \le 2$ this
allows for the investigation of Regge trajectories in this approach. We exemplify such a study
for ground and excited states of light unflavored mesons as well as strange mesons within the 
rainbow-ladder approximation. For the $\rho$- and $\phi$-meson we find a linear Regge trajectory
within numerical accuracy. Discrepancies with experiment in other channels highlight the need 
to go beyond rainbow-ladder and to consider effects such as state mixing and more sophisticated 
quark-antiquark interaction kernels. 
}


\maketitle

\section{Introduction}
Understanding the formation and the structure of hadronic bound-states is one of the most interesting 
-- and difficult -- tasks within QCD. In any gauge-fixed approach to QCD it involves the charting of 
the underlying non-perturbative interactions between quarks and gluon and requires an understanding of
the associated phenomena of dynamical quark mass generation and confinement.
The simplest color neutral state of QCD is the meson, consisting of a quark and an antiquark,
which gives rise to particular combinations of quantum numbers $J^{PC}$ often characterized within 
the quark model. However, similar (and exotic) quantum numbers may arise for so-called hybrid states 
that contain one or more constituent gluons, as well as more complex ones such as glueballs, 
meson molecules and tetraquarks. These states may mix into each other, thus providing a rich and 
complicated spectrum explored in many experiments.

This may be particularly true for the light meson sector, where a huge amount of literature is available 
dealing with this problem. Relativistic quark models, effective chiral Lagrangians, Hamiltonian
approaches, QCD sum rules, Dyson-Schwinger and functional renormalisation group methods as well as 
lattice QCD are methods of choice, see \emph{e.g.}~\cite{Brambilla:2014aaa} for a recent review and a guide 
to further reading. In this work we concentrate on the functional approach via Dyson--Schwinger 
equations (DSEs) and Bethe--Salpeter equations (BSEs), which offers the above-mentioned direct 
connection between the details of the non-perturbative quark-gluon interaction and the relativistic and 
field-theoretical description of bound-states. 

The purpose of this work is twofold. On the one hand, we report on an important technical extension: 
to the well-known representations of (pseudo-)scalar, {(axial-)}\linebreak vector and (pseudo-)tensor states 
\cite{Joos:1962qq,Weinberg:1964cn,Zemach:1968zz,Krassnigg:2010mh} we add an explicit basis construction for mesons with $J=3$. 
This allows, for the first time, the explicit study of Regge-trajectories 
in the DSE/BSE framework. For the light meson sector this may be especially interesting, since the 
conventional picture of a linear rising potential associated with a flux tube that underlies intuitive 
explanations of linear Regge-type behavior hardly seems appropriate. This is, however, the mechanism 
that is built into relativistic quark models relying on linear rising (quasi-)potentials, see 
\emph{e.g.}~\cite{Godfrey:1985xj,Ebert:2009ub}. In contrast, an approach like the DSE/BSE framework 
offers the opportunity to explore alternative mechanisms for the generation of Regge-type behavior 
from the underlying quark-gluon interaction. 

On the other hand, we explore the spectrum of ground and excited states of light mesons with total 
angular momentum $J=0,1,2,3$ starting from the simplest of all truncations, namely a rainbow-ladder 
framework with a flavor-diagonal interaction. It has the merit of preserving chiral symmetry in the 
form of the axial-vector Ward-Takahashi identity thus reproducing important QCD constraints such as
the (pseudo-)Goldstone boson nature of the pseudoscalar mesons and the associated Gell-Mann--Oakes--Renner
relation. However, it has been noted previously
(see \emph{e.g.} Refs.~\cite{Qin:2011xq,Blank:2011ha} and Refs. therein) that apart from the ground 
state pseudoscalar 
and vector channels, this truncation may have serious shortcomings that prevent close quantitative 
contact with experiment. In this work we explore this issue for a wide range of different channels 
and confirm previous work. However, including the $J=3$ results we are able to identify a larger 
pattern that leads to different conclusions than those made in \cite{Qin:2011xq} regarding the 
precise origin of the shortcoming of the rainbow-ladder truncation. This is detailed below.

The paper is organized as follows. In section~\ref{sec:framework} we introduce the framework of the 
DSEs and BSEs, together with a discussion of the Rainbow-Ladder truncation and the model interaction 
employed. In section~\ref{sec:bethesalpeteramplitude} we present the covariant decomposition of 
the Bethe--Salpeter amplitude for $J=0,1,2$, as well as $J=3$. Our numerical methods are discussed 
in section~\ref{sec:numericalmethods} with results given in section \ref{sec:results}. We conclude 
in section~\ref{sec:conclusions}.

\section{Framework}\label{sec:framework}
We work in Euclidean space with the dressed one-particle irreducible Green's functions
of QCD. These are obtained through solutions of their corresponding DSEs, employing
truncations that are designed to maintain important symmetries of QCD such as chiral 
symmetry. The resulting Green's functions serve as input into the BSE for bound
states of a quark and an anti-quark. In the following we summarize the corresponding 
formalism and give details on the approximation scheme used.

\subsection{Quark propagator}\label{sec:quarkpropagator}
\begin{figure}[t]
\begin{center}
\includegraphics[width=0.99\columnwidth]{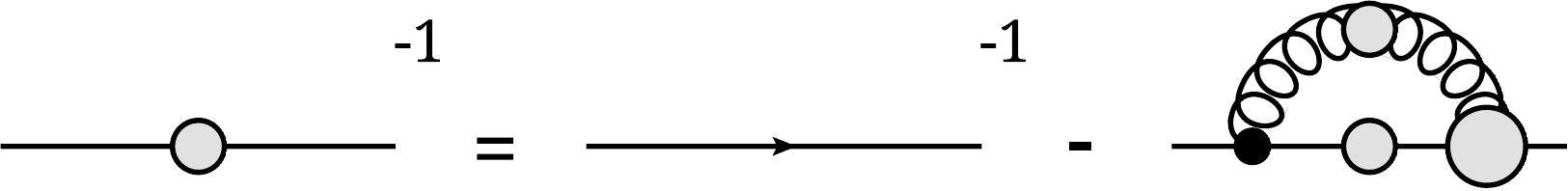}
\caption{The Dyson--Schwinger equation for the fully dressed quark
propagator. Wiggly lines represent gluons and straight lines quarks. Large
filled circles indicate the quantity is fully-dressed, otherwise it is bare.}\label{fig:quarkdse}
\end{center}
\end{figure}

The dressed quark propagator is given by 
\begin{align}
S^{-1}(p) = Z_f^{-1}(p^2)\left( i\pslash + M(p^2)\right)\;,
\end{align}
where the quark wave function is $Z_f(p^2)$ and its mass function $M(p^2)$. The bare quark
propagator is obtained by setting $Z_f(p^2)=1$ and $M(p^2)=m_0$, with the bare quark mass 
$m_0$ related to the renormalized quark mass $m_q$ via $Z_2 \,m_0=Z_2\,Z_m\, m_q$ via the
renormalization factors $Z_2$ and $Z_m$. The dressing functions $Z_f(p^2)$ and $M(p^2)$ are 
obtained as a solution of the quark DSE
\begin{align}\label{eqn:quarkdse}
S^{-1}(p) = Z_2S_0^{-1}+g^2 Z_{1f}C_F\int_k \gamma^\mu S(k)\Gamma^\nu(k,p)D_{\mu\nu}(q)\;,
\end{align}
given pictorially in Fig.~\ref{fig:quarkdse}. Here we use the abbreviation $\int_k=\int d^4k/(2\pi)^4$
and the momentum routing $q=k-p$. The gluons propagator is denoted by $D_{\mu\nu}(q)$ and 
$\Gamma^\nu(k,p)$ the dressed quark-gluon vertex. The corresponding renormalization factor
is $Z_{1f}$. The Casimir factor $C_F=4/3$ stems from the color trace and $g$ is the renormalized coupling of QCD.

We work in Landau gauge, where the gluon propagator is purely transverse and given by 
\begin{align}
D_{\mu\nu}(q) = \left(\delta_{\mu\nu}-\frac{q_\mu q_\nu}{q^2}\right)\frac{Z(q^2)}{q^2}= 
T_{\mu\nu}^q\frac{Z(q^2)}{q^2}\;,
\end{align}
with transverse projector $T_{\mu\nu}^q$. 

\subsection{Bethe--Salpeter equation}\label{sec:bethesalpeterequation}
\begin{figure}[t]
\begin{center}
\includegraphics[width=0.6\columnwidth]{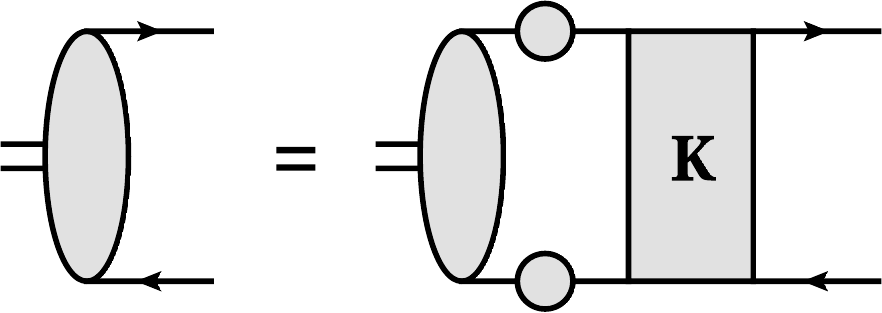}
\caption{The homogeneous Bethe--Salpeter equation for a bound-state of a quark and an antiquark. 
The quark-antiquark interaction kernel $K$ is constrained by requirements of chiral 
symmetry.}\label{fig:mesonbse}
\end{center}
\end{figure}

The Bethe--Salpeter equation, given in Fig.~\ref{fig:mesonbse}, describes a relativistic 
bound-state of mass $M$ calculated through
\begin{equation}
  \left[\Gamma(p;P)\right]_{tu} = \lambda\! \int_k\!\!
  K^{(2)}_{rs;tu}(p,k;P)\left[\chi(k;P)\right]_{sr}\,.
\end{equation}
Here, $\Gamma(p;P)$ is the 
Bethe--Salpeter amplitude, $\chi(k;P)=S(k_+)\Gamma(k;P)S(k_-)$ the corresponding wave function
and $K^{(2)}$ a two-particle irreducible quark anti-quark interaction kernel.
The momenta 
$k_\pm = k + (\xi -1/2 \pm 1/2) P$ 
feature a momentum partitioning parameter 
$\xi$, which has no influence on the bound 
state mass. The BSE is a homogeneous eigenvalue equation 
with a discrete spectrum of solutions at momenta $P^2=-M_i^2$ and eigenvalues $\lambda\left(P_i^2\right)=1$. 
The lightest of these $M_i$ is the ground state solution. 
For a bound state with total angular momentum $J$, both $\Gamma(p;P)$ and $\chi(k;P)$ have $J$ 
Lorentz indices. Their covariant decomposition is a combination of the Dirac representation for 
two composite spin-$1/2$ fermions and an angular momentum tensor.

\subsection{Rainbow-Ladder}\label{sec:rainbowladder}

The essential input into the DSE for the quark propagator and the interaction kernel 
of the two-body Bethe-Salpeter equation is the dressed gluon propagator and the dressed 
quark-gluon vertex. Both the quark self energy and the interaction kernel $K^{(2)}$ are related by
the axial Ward-Takahashi identity (axWTI). Any meaningful approximation of these quantities in
the light quark sector has to satisfy this identity, otherwise essential QCD properties of
dynamical chiral symmetry-breaking such as the (pseudo-)Goldstone nature of the pion and 
the Gell-Mann-Oakes-Renner relation are lost. The simplest construction principle to
satisfy the axWTI is the rainbow-ladder approximation, which is tantamount to taking into
account only the $\gamma_\mu$-structure of the dressed quark-vertex and combining all
dressing effects of the gluon and the vertex into an effective running coupling. The
rainbow-ladder approximation is simple to use and we will therefore employ it in this 
exploratory study. Its merits and deficiencies have been indicated already in the 
introduction and will be discussed in more detail in the results section. 
 
For completeness let us mention, that there have been many efforts to go beyond 
rainbow-ladder. One promising route is to use explicit diagrammatic approximations to 
the DSE of the quark-gluon vertex \cite{Bender:1996bb,Watson:2004kd,Bhagwat:2004hn,%
Matevosyan:2006bk,Alkofer:2008tt,Fischer:2007ze,Fischer:2009jm,Fischer:2008wy}.
This allows the explicit study of the effects of the gluon self-interaction
\cite{Fischer:2009jm} as well as pion cloud effects \cite{Fischer:2008wy} on 
the spectrum of light mesons and baryons \cite{Sanchis-Alepuz:2014wea}. Another 
promising approach uses explicit representations of selected tensor structures of the 
quark-gluon vertex beyond the leading $\gamma_\mu$ piece 
\cite{Chang:2009zb,Chang:2010hb,Chang:2011ei,Heupel:2014ina}. As will be clear from 
the results section it is mandatory to repeat the meson survey performed here in
one or more of the above approaches in future work.

In rainbow-ladder approximation the relevant parts in the quark self-energy simplify 
according to 
\begin{align}
Z_{1f} C_F\frac{g^2}{4\pi} D_{\mu\nu}(q)\Gamma^\nu(k,p) = Z_2^2C_FT_{\mu\nu}^q 
\frac{\alpha_{\mathrm{eff}}(q^2)}{q^2}\gamma^\nu\;,
\end{align}
where we also collected together some numerical constants and the color traces. 
The corresponding two-body kernel, consistent with chiral symmetry, is given by
\begin{align}
K^{(2)}_{rs;tu}(p,k;P) = 4\pi Z_2^2C_F \frac{\alpha_{\mathrm{eff}}(q^2)}{q^2} 
T_{\mu\nu}^q [\gamma^\mu]_{rt}[\gamma^\nu]_{su}\;,
\end{align}
and is a function of $q=k-p$ only. 
In particular, we take the model interaction of Maris and 
Tandy~\cite{Maris:1999nt}
\begin{align}
\alpha_{\mathrm{eff}}(q^2)=\pi\eta^7x^2e^{-\eta^2x}
+\frac{2\pi\gamma_m\left(1-e^{-y}\right)}{\log\left[e^2-1+(1+z)^2\right]}\;,
\end{align}
where $x=q^2/\Lambda^2$, $y=q^2/\Lambda_t^2$, $z=q^2/\Lambda_{\mathrm{QCD}}^2$. 
Here $\Lambda_t=1$~GeV is a regularization parameter for the perturbative logarithm;
its value has no material impact on the numerical results. The QCD-scale 
$\Lambda_{\mathrm{QCD}}=0.234$~GeV controls the running of the logarithm with
anomalous dimension $\gamma_m=12/25$ corresponding to four active quark flavors.
The infrared strength of this model is controlled by the parameters $\Lambda$ and 
$\eta$. While $\Lambda = 0.72$ GeV is fixed from the pion decay constant, there is considerable
freedom to vary the dimensionless parameter $\eta$. We will use ${\eta=1.8\pm0.2}$ for 
which ground-state observables are insensitive and discuss the effects of this variation
on the excited states. 

\section{Covariant Bethe-Salpeter amplitude}\label{sec:bethesalpeteramplitude}
It is well known that composite states of particles in the $\left(j,0\right)\oplus\left(0,j\right)$-representation can be constructed 
by forming direct products of the particle's representation~\cite{Joos:1962qq,Weinberg:1964cn}. For fermions, $j=1/2$, this reduces
to the Dirac spinor formalism and thus is given by the usual Dirac matrices.

For a meson in the rest frame with center-of-mass momentum $t_\mu$ and
relative quark momentum $r_\mu$, grouped by their transformation under parity we have
\begin{align}\label{eqn:scalarinvariants1}
D^{(\ONE)}&=
\left(\begin{array}{ccccc}
\;\ONE\; &
\phantom{\gamma_5}t_\mu \gamma^\mu\;&
\phantom{\gamma_5}r_\mu \gamma^\mu\;&
\phantom{\gamma_5}r_\mu t_\nu \frac{1}{2}\left[\gamma^\mu,\gamma^\nu\right]
\end{array}\right)\;, \\
D^{(5)}&=
\left(\begin{array}{ccccc}
\gamma_5\;&
\gamma_5 t_\mu \gamma^\mu\;&
\gamma_5 r_\mu \gamma^\mu\;&
\gamma_5 r_\mu t_\nu \frac{1}{2}\left[\gamma^\mu,\gamma^\nu\right]
\end{array}\right)\;,
\end{align}
for scalar, $D^{(\ONE)}$, and pseudoscalar, $D^{(5)}$, invariants respectively. Thus, for a 
bound-state of two fermions with definite parity, the basic number of scalar invariants equals
four. Furthermore, it is convenient to replace the relative momentum $r_\mu$ by
\begin{align}\label{eqn:transverseQ}
Q_\mu = \tau^{(t)}_{\mu\nu}\;r^\nu\;,
\end{align}
where $ \tau^{(t)}_{\mu\nu}=\delta_{\mu\nu}-t_\mu t_\nu/t^2$ is a transverse projector.
Then, appropriate scalar and pseudoscalar invariants are
\begin{align}\label{eqn:scalarinvariants2}
\bar{D}^{(\ONE)}&=
\left(\begin{array}{ccccc}
\;\ONE\; &
\phantom{\gamma_5}\slashed{t}\;&
\phantom{\gamma_5}\slashed{Q}\;&
\phantom{\gamma_5}\slashed{Q}\slashed{t}
\end{array}\right)\;,\;\;\;\;
\bar{D}^{(5)}=\gamma_5\bar{D}^{(\ONE)}\;,
\end{align}
which simplifies the operation of charge conjugation due to the property that $Q\cdot t=0$.

Then, a bound state with zero total angular momentum and definite parity $P$ is decomposed in terms of 
four components
\begin{align}
\Gamma^{(P)}(r,t) &= \sum_{i=1}^{4} \left[\lambda_i  \bar{D}_{i}^{(P)}\right]\;.
\end{align}

For non-zero total angular momentum $J$, the scalar invariants
must be coupled with an angular momentum tensor. This rank $J$ tensor, $T_{a_1,\ldots a_J}$, has 
$2J+1$ independent components in three spatial dimensions, corresponding to the possible spin 
polarisations~\cite{Zemach:1968zz}. This tensor must be 
symmetric in all indices and traceless with respect to contraction of any pair of indices.
This generalizes to $3+1$ dimensions by imposing transversality of each 
index with respect to the total momentum.

Thus, to obtain a tensor corresponding to total angular momentum $J$, we construct the 
symmetric $J$-fold tensor product of a transversal projector transforming like a vector, and 
subtract traces with respect to every pair of indices. The case $J=1$ will provide tensors that 
form the building blocks for states of higher total angular momentum.

Then, in general a meson of spin $J>0$ and parity $P$ has eight components and is written
\begin{align}
\Gamma^{(P)}_{\mu_1\ldots\mu_J}(r,t) &= \sum_{i=1}^{4} \left[\lambda_i Q_{\mu_1\ldots\mu_J} \bar{D}_{i}^{(P)}+\lambda_{i+4} T_{\mu_1\ldots\mu_J} \bar{D}_{i}^{(P)}\right]\;,
\end{align}
where the $Q_{\mu_1\ldots\mu_J}$, $T_{\mu_1\ldots\mu_J}$ are defined below and $\lambda_i = \lambda_i(r,t)$.

\subsection{Total angular momentum $J=1$}

For the case of $J=1$ we can immediately write down the two rank 1 tensors for a bound state of two
fermions: they are the transversely projected quantities $Q_\mu$ and $T_\mu$ defined 
\begin{align}\label{eqn:transverseprojectors}
Q_{\mu} = \tau^{(t)}_{\mu\nu}\; r^\nu\;,\qquad
T_{\mu} = \tau^{(t)}_{\mu\alpha}\; \tau^{(Q)}_{\alpha\nu}\gamma^\nu\;.
\end{align}
Here $Q$ is the same quantity as defined in Eq.~\eqref{eqn:transverseQ} and we introduced the 
additional transverse projector $\tau^{(Q)}_{\alpha\nu}$ so that the resulting basis is 
conveniently orthogonal. The explicit components of this basis can be found {\it e.g.} in 
Ref.~\cite{Maris:1999nt}.

\subsection{Total angular momentum $J=2$}
For total angular momentum $J=2$ we construct the $2$-fold tensor products of $Q_{\mu_i}$ and 
$T_{\mu_i}$. Since the product of two or more $T_{\mu_i}$ is degenerate, this gives 
\begin{align}
  \tilde{Q}_{\mu_1\mu_2} &= Q_{\mu_1}Q_{\mu_2} \; , \\
  \tilde{T}_{\mu_1\mu_2} &=T_{(\mu_1}Q_{\mu_2)} \;,
\end{align}
where ${(\ldots)}$ denotes the symmetrization of the indices without normalization 
$\nicefrac{1}{J!}$.
To satisfy the criteria of being angular momentum tensors we then subtract the trace-part to give~\cite{LlewellynSmith:1969az,Krassnigg:2010mh}
\begin{align}
  Q_{\mu_1\mu_2} &= Q_{\mu_1} Q_{\mu_2} - \frac{1}{3}Q^2 \tau_{\mu_1\mu_2} \;, \\
  T_{\mu_1\mu_2} &= T_{(\mu_1} Q_{\mu_2)} \;.
\end{align}
The explicit components of this basis can be found {\it e.g.} in Ref.~\cite{Krassnigg:2010mh}.

\subsection{Total angular momentum $J=3$}
For total angular momentum $J=3$ we construct the $3$-fold tensor products of $Q_{\mu_i}$ and 
$T_{\mu_i}$
\begin{align}
  \tilde{Q}_{\mu_1\mu_2\mu_3} &= Q_{\mu_1}Q_{\mu_2}Q_{\mu_3} \; , \\
  \tilde{T}_{\mu_1\mu_2\mu_3} &= T_{(\mu_1}Q_{\mu_2}Q_{\mu_3)}\;.
\end{align}
To satisfy the requirements of angular momentum tensors we subtract the trace part, yielding
\begin{align}
  Q_{\mu_1\mu_2\mu_3} &= \tilde{Q}_{\mu_1\mu_2\mu_3} - \frac{1}{5} 
   \tau_{(\mu_1\mu_2}\tilde{Q}^{\kappa\kappa}_{\phantom{\kappa\kappa}\mu_3)}  \;\; , \nonumber\\
  &= Q_{\mu_1}Q_{\mu_2}Q_{\mu_3} - \frac{ 1 }{5}Q^2 
   \tau_{(\mu_1\mu_2} Q_{\mu_3)} \; \;, \\
  T_{\mu_1\mu_2\mu_3} &=\tilde{T}_{\mu_1\mu_2\mu_3} -\frac{1}{5} 
   \tau_{(\mu_1\mu_2}\tilde{T}^{\kappa\kappa}_{\phantom{\kappa\kappa}\mu_3)}  \nonumber \\
  &= T_{(\mu_1}Q_{\mu_2} Q_{\mu_3)} - \frac{1}{5} Q^2\tau_{(\mu_1\mu_2} T_{\mu_3)}   \;,
\end{align}
which has not been explored in this approach before. The explicit representation of this basis is given by
\begin{align}
\Gamma^{(\ONE)}_{\mu_1\mu_2\mu_3}(r,t) &= 
   Q_{\mu_1\mu_2\mu_3} \left[ \lambda_1 \ONE + \lambda_2\slashed{t} +\lambda_3 \slashed{Q} + \lambda_4\slashed{Q}\slashed{t}  \right]\nonumber\\
 &+T_{\mu_1\mu_2\mu_3}\; \left[ \lambda_5 \ONE + \lambda_6\slashed{t} +\lambda_7 \slashed{Q} + \lambda_8\slashed{Q}\slashed{t}  \right]\;,
\end{align}
with $\lambda_i=\lambda_i(r,t)$ scalar coefficients. Multiplying through by $\gamma_5$ would yield the 
$\Gamma^{(5)}_{\mu_1\mu_2\mu_3}(r,t)$ basis decomposition.

\section{Numerical Methods}\label{sec:numericalmethods}
Here we give a brief summary of the numerical methods used for this work. Primarily, this concerns
the solution of non-linear integral equations at complex Euclidean momenta, followed by finding
eigenvalues and the corresponding eigenvectors of a linear system in matrix form. We also discuss the 
means by which the higher mass states are obtained.

\subsection{Quark propagator for complex momenta}
\begin{figure}[!b]
\begin{center}
\includegraphics[width=0.8\columnwidth]{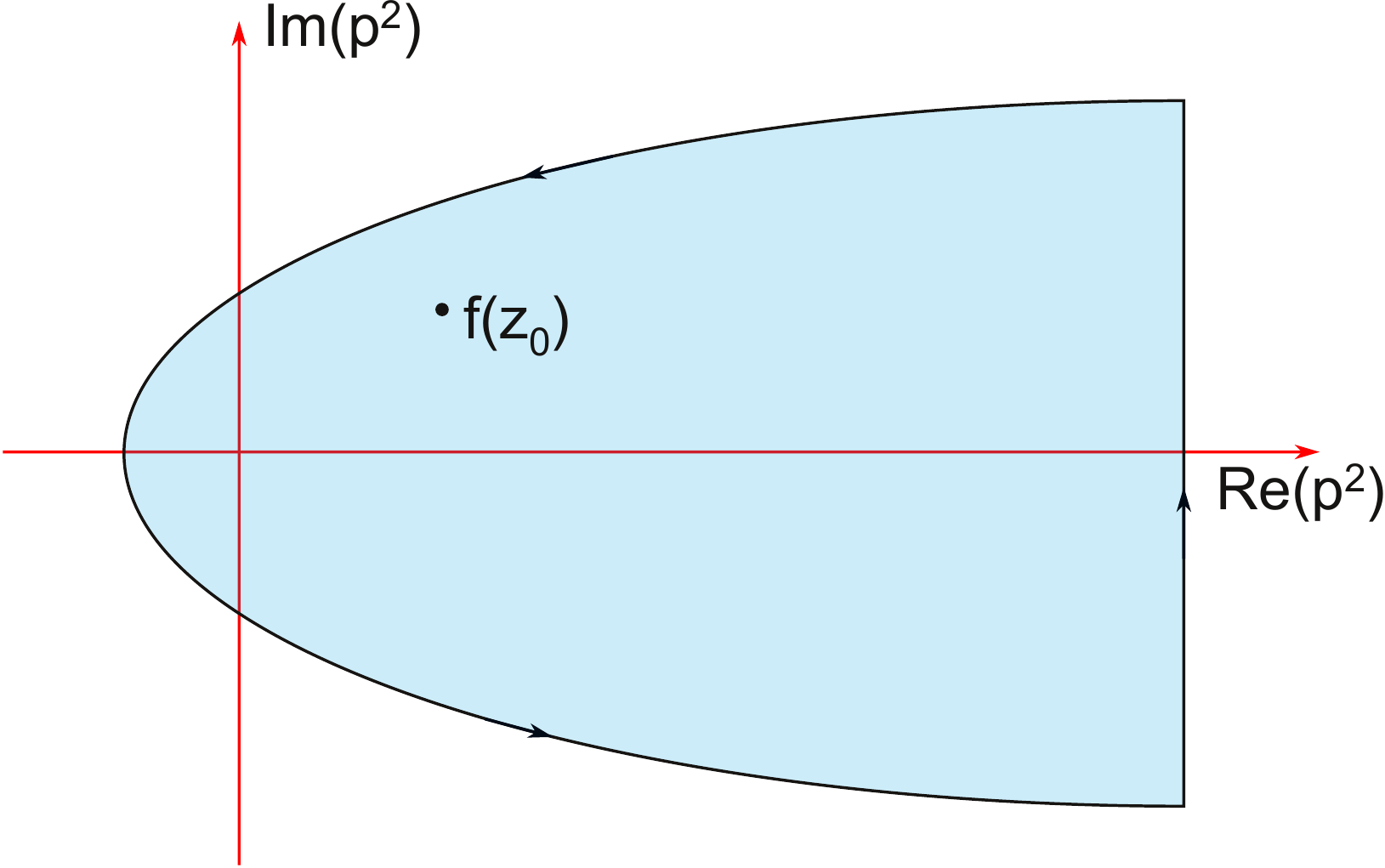}
\caption{Sketch of the integration contour for the determination of the quark propagator in
the complex plane.}\label{fig:contour}
\end{center}
\end{figure}
In the BSE due to the external total momentum of the bound state one needs to evaluate the
internal propagators on the right hand side in a parabola region sketched in Fig.~\ref{fig:contour}.
The quark propagator at these complex momenta $p^2$ could be evaluated directly from its 
DSE, Eq.~\eqref{eqn:quarkdse}, provided one knows the quark propagator for spacelike Euclidean 
momenta $p^2>0$ and the quark-gluon vertex as well as the gluon propagator for complex (gluon) 
momenta. In general, however, this is not the case and one has to rely upon numerical input for
the gluon propagator or the vertex. In order to make our procedure sufficiently general for 
later studies, we use an alternative strategy.  We 
change the momentum routing in the DSE such that the external complex-valued momentum flows 
through the internal quark propagator. This complex shift in the quark momentum entails that 
a parabolic region of the complex plane is probed by the internal quark, similar to 
that in the BSE, see Fig.~\ref{fig:contour}. The DSE is then solved iteratively either on a 
momentum grid inside the parabolic region \cite{Fischer:2005en} or on the boundary supplemented
with Cauchy's theorem. Here, we use the latter method: given a function $f(z)$ defined on the 
boundary of a closed contour $z\in\mathcal{C}$, we have for any $z_0$ inside 
\begin{align}\label{eqn:cauchy}
f(z_0) = \frac{1}{2\pi i} \oint_{\mathcal{C}} \frac{dz f(z)}{z-z_0}\simeq \frac{1}{2\pi i} \sum_{i} \frac{w_i f(z_i)}{z_i-z_0}\;,
\end{align}
where the integral has been approximated by some quadrature formula with weights $w_j$ and 
abscissa $z_j$. This is paired with a parametric mapping that describes the contour's boundary.
Numerically this procedure poses a challenge when $z_0$ approaches the abscissa $z_i$. This can be 
mitigated through the use of the barycentric formula~\cite{Berrut_barycentriclagrange}
\begin{align}\label{eqn:cauchybary}
f(z_0) = \frac{\sum_{i} \bar{w}_i f(z_i)  }{\sum_{i} \bar{w}_i }\;,\;\;\;\;\bar{w}_i=w_i /\left(z_i-z_0\right)\;.
\end{align}
If the contour $\mathcal{C}$ is such that it encounters complex conjugate poles in the quark propagator, 
Eqs.~\eqref{eqn:cauchy}--\eqref{eqn:cauchybary} can be modified to include the residues. However, 
it is still a technical challenge to determine and include such poles numerically in a non-linear integral
equation such as the quark DSE. 

\subsection{Calculating bound state masses}\label{sec:boundstates}
The Bethe--Salpeter equation is reduced to an eigenvalue problem for $\Gamma=\lambda M\cdot \Gamma$.
The amplitude $\Gamma=\Gamma(p;P)$, for total momentum $P$, is a function of the relative quark momentum $p$
and the angle $\widehat{p\cdot P}$. This angular dependence is expanded as a sum of Chebyshev polynomials, 
reducing the system to a coupled system of linear equations in one variable, $p^2$.

The matrix $M$ represents the coupling of this amplitude to the interaction kernel $K$ and its subsequent integration. 
It is solved as an eigenvalue equation using the \emph{Eigen} library~\cite{eigenweb}. We specify the $J^P$ of the 
state through the choice of the covariant decomposition, section~\ref{sec:bethesalpeteramplitude}, and determine the 
$C$-parity of the state by examining the symmetry properties of the eigenvector. 
Excited states are obtained by finding solutions $\lambda=1$ higher in the mass spectrum.

Since excited states, and those with $J>2$, are typically heavy we find ourselves in the position that
the parabolic region in the complex plane for which the quark propagator has been calculated is too small. The
region cannot, at present, be extended due to the presence of propagator poles that must be taken into account
self-consistently. A similar problem is encountered when one attempts to calculate the mass spectrum of heavy-light
mesons; this is a general problem for Euclidean bound-state calculations in general within the Bethe--Salpeter framework, see \cite{Chang:2013nia,Dorkin:2013rsa} for recent attempts to 
circumvent this problem.

Here, we pursue two approximations to gather information beyond this constraint. The first is to employ
the Cauchy theorem for $z_0$ outside of the contour. Of course, this is only an approximate analytic 
continuation which, however, while not mathematically precise nevertheless culminates in well-behaved 
quark dressing functions. 

\begin{figure}[t]
\begin{center}
\includegraphics[width=1.0\columnwidth]{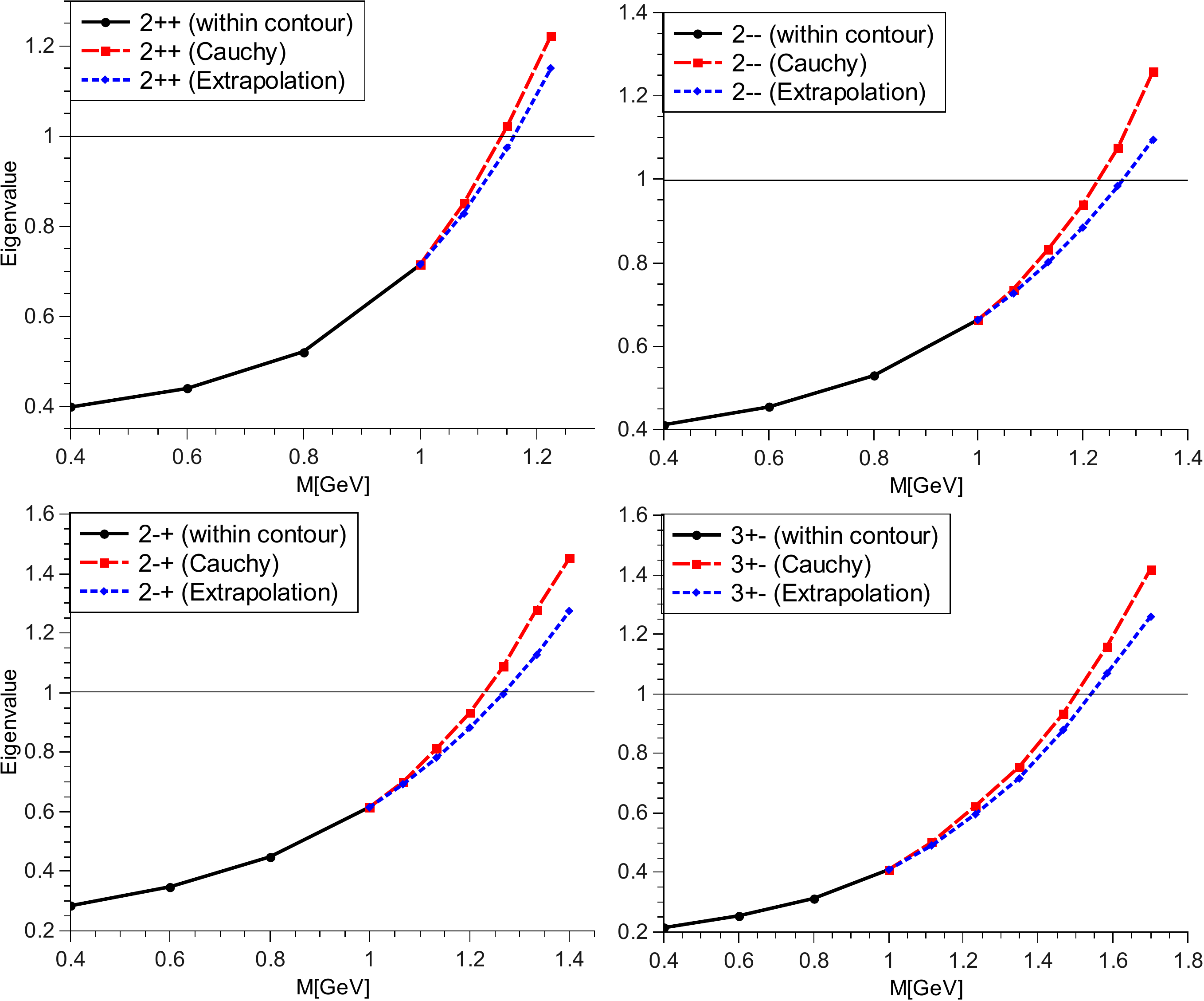}
\caption{Comparison of the eigenvalue curves obtained by employing the Cauchy integral formula for outside of the contour and the barycentric rational interpolation. 
The black curve represents the eigenvalues at masses obtained within contour. }\label{fig:extrapolation}
\end{center}
\end{figure}

\begin{figure*}[tp]
\begin{center}
\includegraphics[width=0.95\textwidth]{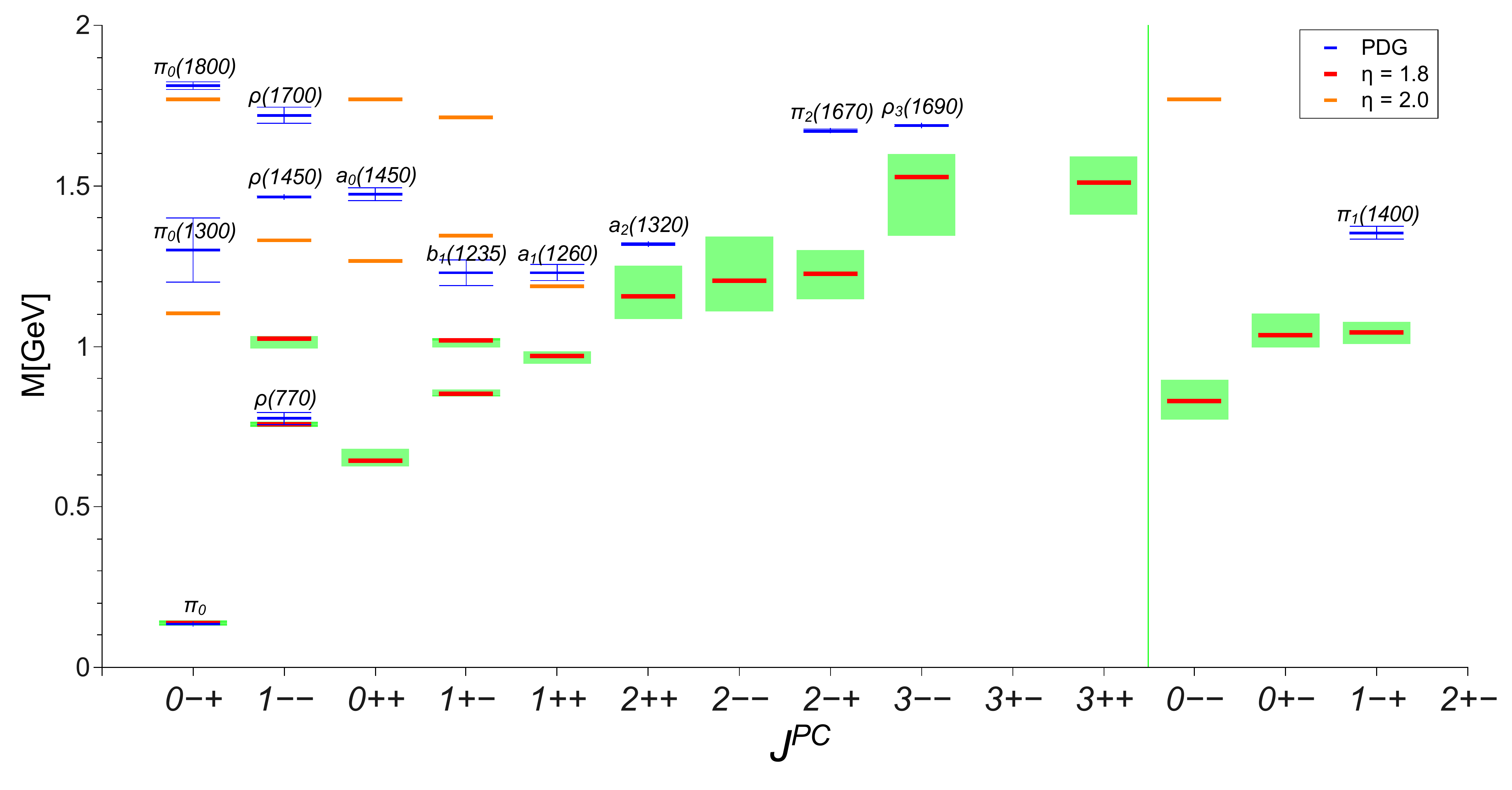}
\caption{(color online) (top) The calculated $n\bar{n}$ spectrum, compared to the isovector mesons as measured in 
experiment. The green bands correspond to the 
variation $\eta=1.8\pm0.2$. Due to the structure of the propagator, in the case of $\eta=2.0$ more states are accessible; 
these are given by the single orange lines. The states to the right of the dividing line correspond to exotic quantum 
numbers.}\label{fig:spectrumnn}
\end{center}
\end{figure*}

Alternatively one can extrapolate the eigenvalue according to~\cite{Blank:2011ha},
where it was shown that linear extrapolation in $\lambda_i^{-1}(P^2)-1$ is reliable for the pseudoscalar excited states; however this does not hold for the other channels. Therefore we used barycentric rational interpolation
\begin{align}
R(x) = \frac{\sum^{N-1}_{i=0} \frac{w_i}{x-x_i}y_i }{\sum^{N-1}_{i=0} \frac{w_i}{x-x_i} }\;,
\end{align}
where $w_i$ are the weights at desired order $d$ are given by
\begin{align}
w_k = \sum^{k}_{i=k-d}(-1)^k \prod^{i+d}_{j=i,j \neq k} \frac{1}{x_k-x_j}\;.
\end{align}
The comparison of these two techniques is shown in Fig.~\ref{fig:extrapolation}. As can be seen, 
the masses of the bound states deviate within $5-10\%$ of the mean value with an overall tendency 
that the Cauchy integral formula gives a lower bound. For the purpose of the exploratory study 
reported here, we regard this accuracy as sufficient.

\section{Results}\label{sec:results}
\begin{table}[!b]
\renewcommand{\arraystretch}{1.3}
\caption{Allowed quantum numbers for a neutral $q\bar{q}$ state in the quark model. }
\label{tab:qmodel}
\begin{small}
\setlength{\tabcolsep}{3.5pt}
\begin{tabular}{lll|lll|lll|lll|lll}
\hline\noalign{\smallskip}
L & S & $J^{PC}$ & L & S & $J^{PC}$ & L & S & $J^{PC}$ & L & S & $J^{PC}$ & L & S & $J^{PC}$ \\
\noalign{\smallskip}\hline\noalign{\smallskip}
0 & 0 & $0^{-+}$ & 1 & 0 & $1^{+-}$ & 2 & 0 & $2^{-+}$ & 3 & 0 & $3^{+-}$ & 4 & 0 & $4^{-+}$ \\
0 & 1 & $1^{--}$ & 1 & 1 & $0^{++}$ & 2 & 1 & $1^{--}$ & 3 & 1 & $2^{++}$ & 4 & 1 & $3^{--}$ \\
  &   &          & 1 & 1 & $1^{++}$ & 2 & 1 & $2^{--}$ & 3 & 1 & $3^{++}$ & 4 & 1 & $4^{--}$ \\
  &   &          & 1 & 1 & $2^{++}$ & 2 & 1 & $3^{--}$ & 3 & 1 & $4^{++}$ & 4 & 1 & $5^{--}$ \\  
\noalign{\smallskip}\hline
\end{tabular}
\end{small}
\end{table}
The quantum numbers of a meson in the non-relativistic quark model are obtained from the spin, $S$, 
and relative orbital angular momentum $L$ of the $q\bar{q}$ system, which combine to give the total 
spin $J=L\oplus S$. The total parity, $P$, charge parity, $C$, and $G$ parity are given by
\begin{align}
P\left(q\bar{q}\right) &= -(-1)^{L}\;, \\
C\left(q\bar{q}\right) &= \phantom{-}(-1)^{L+S}\;, \\
G\left(q\bar{q}\right) &=\phantom{-}(-1)^{L+S+I}\;,  
\end{align}
where $C$ parity only applies to charge neutral states and is generalized to $G$ parity for isospin $I=1$. 
Thus, the quark model yields the possible $J^{PC}$ quantum numbers in Table~\ref{tab:qmodel}.
This leaves us with five states (for $J\le3$) that are considered exotic:
$J^{PC} = 0^{--}$, $J^{PC} = 0^{+-}$, $J^{PC} = 1^{-+}$, $J^{PC} = 2^{+-}$,  and $J^{PC} = 3^{-+}$.

\begin{figure*}[t]
\begin{center}
\includegraphics[width=0.98\textwidth]{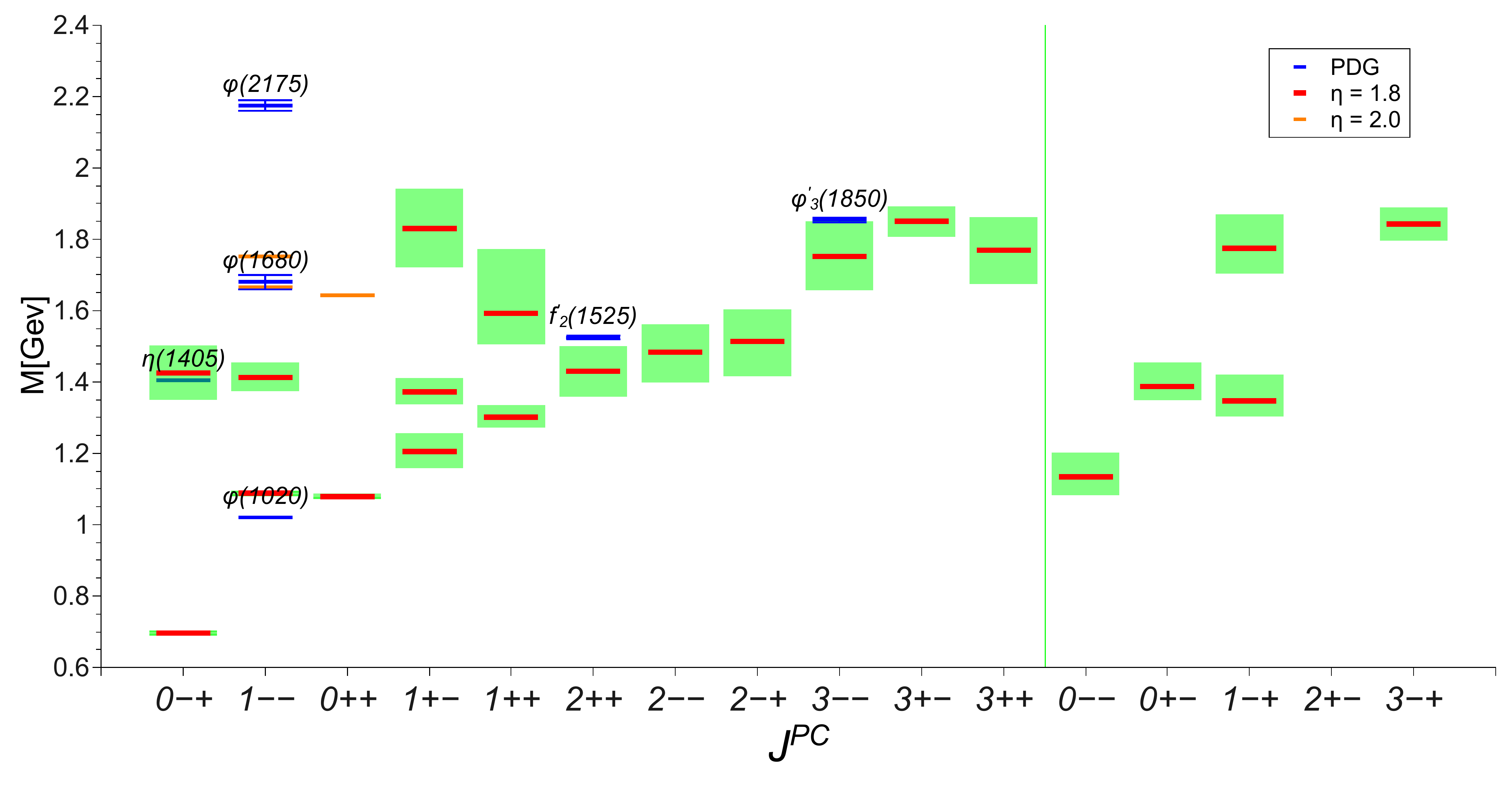}
\caption{(color online) Calculated $s\bar{s}$ spectrum, compared to experiment. The green bands correspond to the 
variation $\eta=1.8\pm0.2$. Due to the structure of the propagator, in the case of $\eta=2.0$ more states are accessible; 
these are given by the single orange lines. The states to the right of the dividing line correspond to exotic quantum 
numbers.}\label{fig:spectrumss}
\end{center}
\end{figure*}

\subsection{Light unflavored mesons}\label{res:light}
In the RL approximation the interaction kernel admits no mixing between states. Furthermore
we work in the isospin symmetric limit using equal current quark masses $m_u=m_d=0.0037$ GeV
at a renormalization scale of $\mu = 19$ GeV. Thus, 
our calculated meson spectrum is degenerate in the isoscalar/isovector channel for $n=u,d$ 
quarks. Th explicit numbers can be found in the Appendix in 
Table \ref{tab:results}. In Fig.~\ref{fig:spectrumnn} we 
display the resulting spectrum for $n\bar{n}$ mesons, and compare with the isovector 
channel from experiment. The input up/down quark masses are fixed such that the experimental 
mass of the $\pi_0$ is reproduced. The resulting ground state mass in the vector channel
is also in good agreement with experiment. This is not true, however, for the scalar and 
axialvector states as noted frequently before, see e.g. \cite{Watson:2004kd}. Here, the
deficiency of the rainbow-ladder truncation is obvious and on the 20-40 \% level. In the
scalar channel there is some evidence that the lowest lying nonet may not be identified
as simple quark-antiquark states, but may be better described as tetraquarks, see {\it e.g.}
\cite{Jaffe:1976ig,Giacosa:2006tf,Ebert:2008id,Parganlija:2012fy,Heupel:2012ua} and Refs. 
therein. Therefore we compare with the $a_0(1450)$, noting that in rainbow-ladder and without
potential mixing with the scalar glueball state there is no hope to reproduce the experimental 
value. The situation is considerably better for the lowest lying tensor state \cite{Krassnigg:2010mh}, 
which for the upper value of the considered $\eta$-band is even on the 5 $\%$ level compared
to the experimental value. While the other tensor states are again far off, at least where
comparison with experiment is possible, the situation is again acceptable for the 
tensor meson with $J=3$ and $PC=\{--\}$. Its mass of $1528^{+71}_{-184}$~MeV
compares well with both the isovector $\rho_3$ of mass $1688.8\pm2.1$~MeV (shown in the figure)
and the isoscalar $\omega_3$ of mass $1667\pm4$~MeV with again a deviation on the 5 $\%$ level
for the upper range of the $\eta$-band. In contrast, we find no bound state in the $J^{PC}=3^{+-}$-channel,
whereas for the $J^{PC}=3^{++}$ state with mass $1510^{+81}_{-100}$~MeV there is no well established 
experimental counterpart.

It is interesting to muse about the difference between the corresponding channels $J^{PC}=0^{-+},1^{--}$
as well as $J^{PC}=1^{--},2^{++},3^{--}$ in good agreement with experiment and the other channels that 
are further off, using
notions of the (pseudo)-potentials in the quark model. In this language, what distinguishes these 
channels from the others is that the non-contact part of the spin-spin interaction 
is vanishing or small: for the hyperfine splitting between the pseudoscalar and vector 
channels the contact part of the spin-spin interaction is dominant, whereas for the $J^{PC}=2^{++},3^{--}$
states the spin-orbit forces prevail. For all other channels considered, there are sizable 
contributions from the tensor part of the spin-spin interaction. Since these are the channels
that are off, we conclude, that the rainbow-ladder interaction roughly reproduces the size of
the contact part of the spin-spin interaction and the spin-orbit force, but materially overestimates
the binding in the tensor part of the spin-spin interaction. Note, that this conclusion is 
different than the one drawn in \cite{Qin:2011xq} based on only a subset of the states considered here.  
We come back to this discussion in section \ref{res:regge}.

As for the exotic channels we find states for $J^{PC}=0^{--},0^{+-}$ with no experimentally established
counterpart, whereas our value for the $J^{PC}=1^{-+}$ is about 25 $\%$ lower than the $\pi_1(1400)$.
This finding is consistent with the ones in the axial-vector channels. In the exotic channels
with $J=2,3$ we do not find bound state.

Finally let us comment on the excited states. These are in general much too low \cite{Holl:2004fr}
in agreement with the general finding for the ground states. A variation of the $\eta$-value
in general does not improve this picture; also it is noteworthy that higher excited states only
appear for very specific values of $\eta$. This suggests a dependency of the excited states on 
the details of the momentum and tensor dependence of the quark-gluon interaction that needs to be explored 
in future work. 

Next we discuss the $s\bar{s}$ spectrum given in the Appendix and displayed in Fig.~\ref{fig:spectrumss}.
Here the input value of the strange quark mass of $m_s(19 \,\mbox{GeV}) = 0.085$ GeV at the renormalization
point is determined from matching to the experimental value of the kaon discussed below.
First note that the pseudoscalar $s\bar{s}$-state is too light in this truncation since neither the 
effect of the $U_A(1)$ anomaly (see {\it e.g.} \cite{Alkofer:2008et} for a treatment of the anomaly
in the BSE formalism) nor flavor mixing with the $n\bar{n}$ states is considered. For the excited 
state in the pseudo-scalar channel the surprisingly excellent agreement with the $\eta(1405)$ extracted from
experiment may be accidental. In the vector channel, where mixing effects do not play a major role we
observe good agreement of our bound state mass with experiment. The same is true for the $J^{PC}=2^{++}$
and $J^{PC}=3^{--}$ channels, where the upper boundary of the $\eta$-band almost reproduces the
experimental values for the $f_2(1525)$ and the $\varphi_3(1850)$. Again, these are the channels
with dominating spin-orbit forces in the language of the potential models. In general, the pattern of
states in the $s\bar{s}$ spectrum is very similar to the one found for the $n\bar{n}$ mesons due
to the flavor independence of the underlying rainbow-ladder interaction model. 

\subsection{Strange Mesons}\label{res:strange}
In the case of strange mesons, $n\bar{s}$, one is no longer able to assign either $C$ or $G$ 
parity to a state. Thus, here there are no states with explicitly exotic quantum numbers.

\begin{figure*}[tp]
\begin{center}
\includegraphics[width=0.98\textwidth]{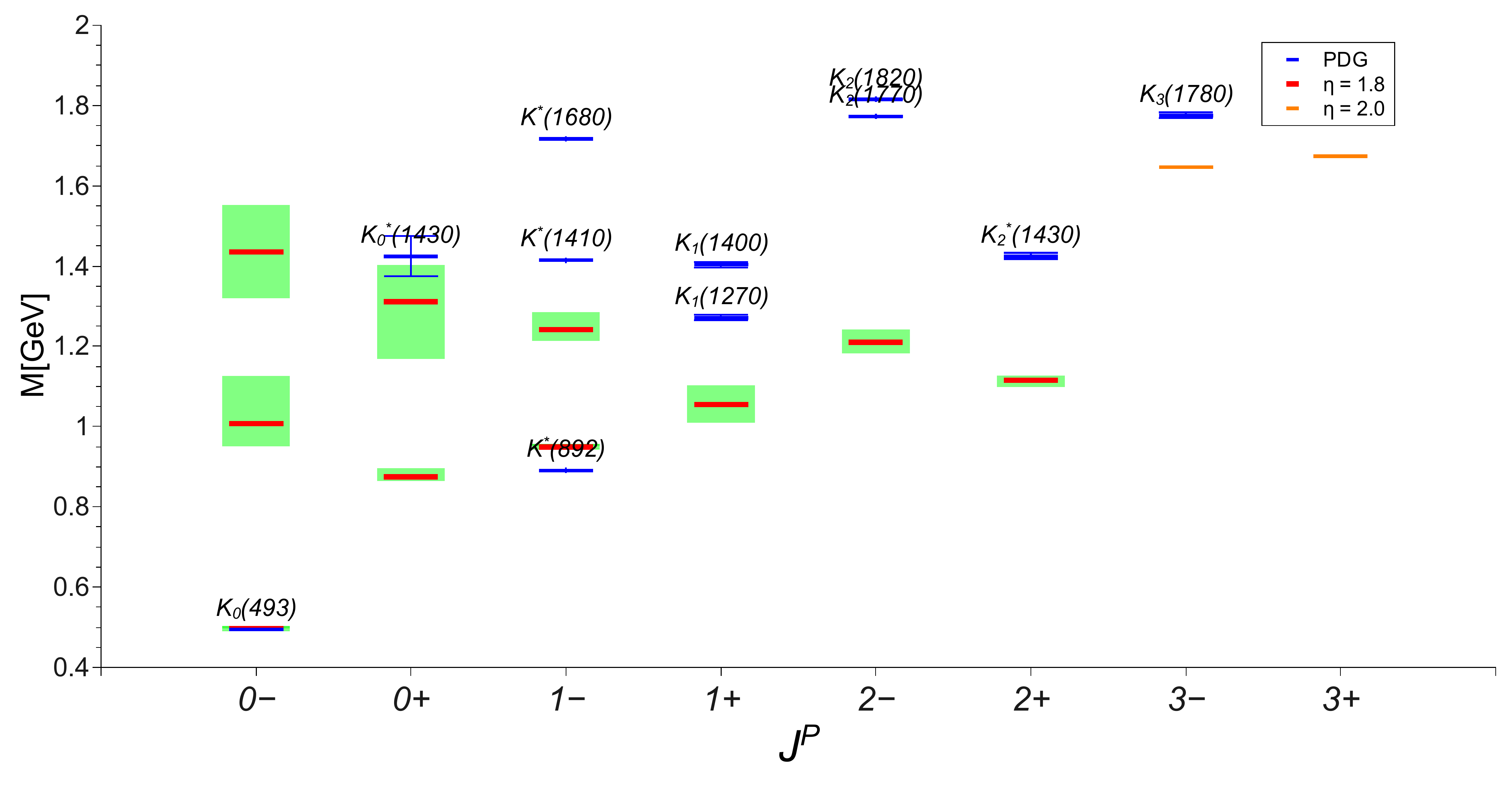}
\caption{(color online) Our calculated $n\bar{s}$ spectrum, compared to experiment. The green bands 
correspond to the variation $\eta=1.8\pm0.2$. Due to the structure of the propagator, in the case 
of $\eta=2.0$ more states are accessible; these are given by the single orange lines. The states 
to the right of the dividing line correspond to exotic quantum numbers.
 }\label{fig:spectrumns}
\end{center}
\end{figure*}

The spectrum, as calculated within the rainbow-ladder approximation, is given in Fig.~\ref{fig:spectrumns}. 
As already mentioned above, the strange quark mass is chosen such that the calculated $K^{0,\pm}$ 
is in agreement in experiment; the remaining spectrum is a result of the model.
While the vector ground state is in reasonable agreement with experiment, the remaining spectrum 
does not fare so well (as in the unflavored case).

Along with the usual $J=1$ and $J=2$ mesons, we find two states with $J=3$, one with positive 
and one with negative parity. For the latter, we have a mass of $1646.9$ (found for $\eta=2.0$ only) 
which compares well with the experimentally known $K_3^\star$ whose mass $1776\pm7$ is within $10\%$. 
The positive parity state is similar in mass, $1673.4$, but the putative $K_3$ has not been seen in 
experiment. 

The results strongly indicate that the $n\bar{s}$ system should be investigated in a beyond rainbow-ladder 
approximation, in order to find stronger agreement for the majority of low-lying states. In particular, the
$2^+$ channel is interesting since the experimentally observed states are considerably higher in mass than 
the calculated ones, in contrast to the findings discussed before in the flavor diagonal channels. On the
other hand, our numerical error in extracting the bound state masses is considerably higher in the
non-diagonal flavor case than in the diagonal one such that it is not clear whether the deficiency is in the
interaction or in our numerical procedure. This needs to be explored further.

\subsection{Regge trajectories}\label{res:regge}

\begin{figure}[!t]
\begin{center}
\includegraphics[width=0.999\columnwidth]{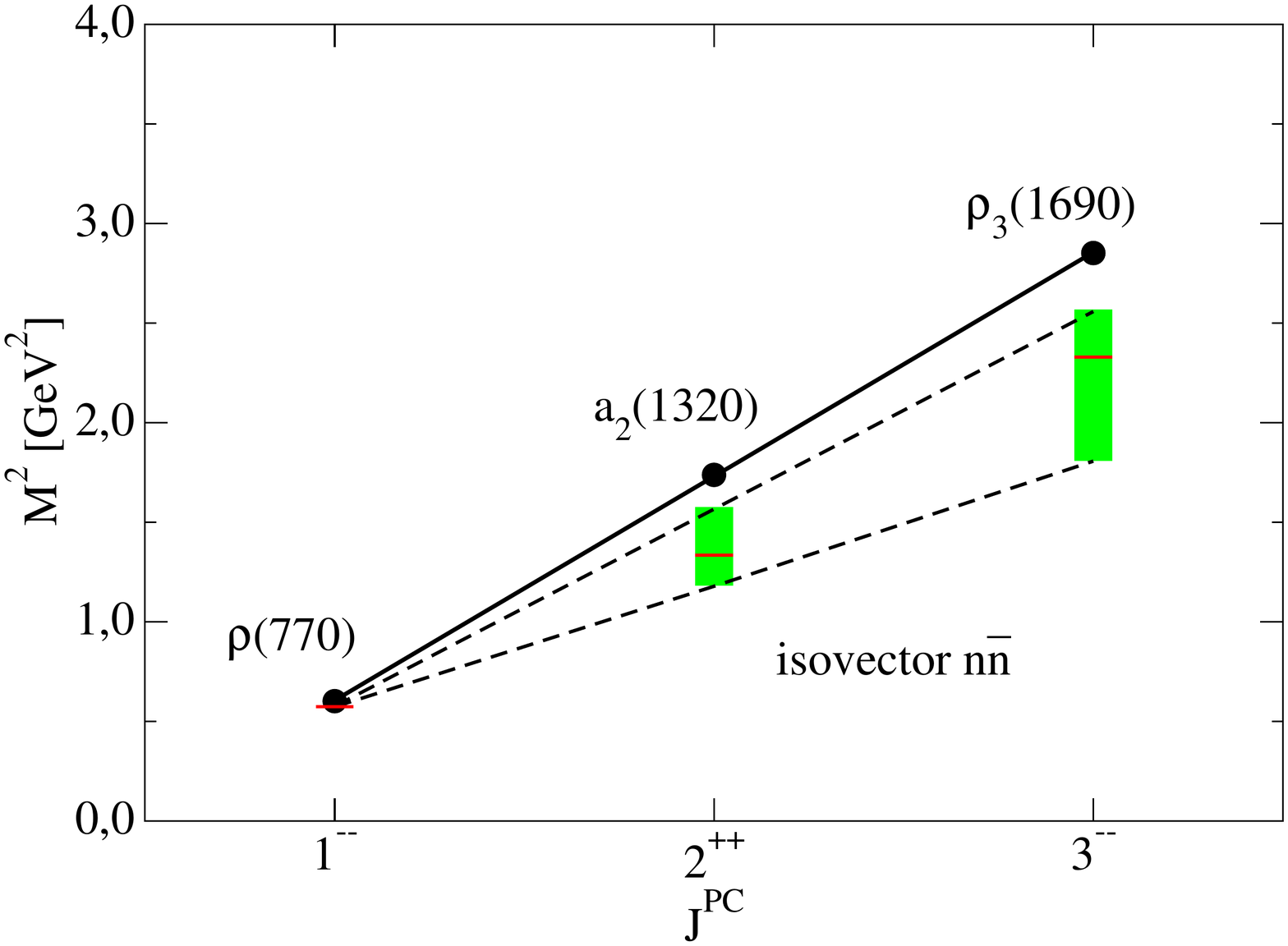}\\
\includegraphics[width=0.999\columnwidth]{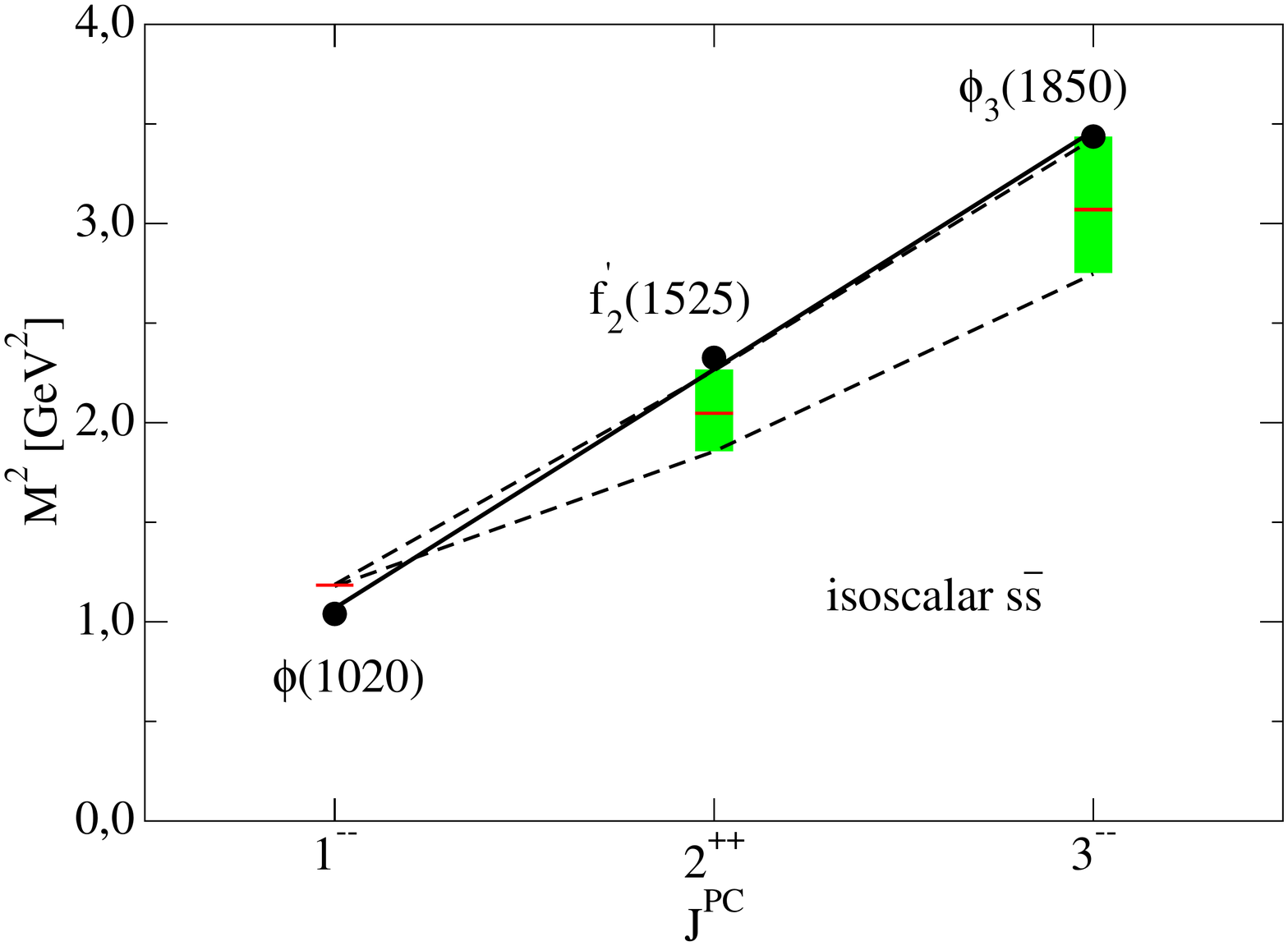}
\caption{Regge trajectories for isovector $n\bar{n}$ (upper plot)
and isoscalar $s\bar{s}$ mesons (lower plot) with natural parity. 
Filled circles correspond to experimental data, while calculated values are given by the red marks
for $\eta = 1.8$ and the green bands for $\eta = 1.8 \pm 0.2$. The resulting Regge trajectories for 
the upper and lower end of the bands are displayed by the dashed lines. Not shown is the numerical
error of our mass extraction procedure, which is of the order of 5-10 $\%$ for the $J=2,3$ states.}\label{fig:regge}
\end{center}
\end{figure}

Finally, we present results for Regge trajectories in Fig.~\ref{fig:regge} for natural parity states. 
We only take into 
account trajectories with at least three states, which leaves the ground state isovector $n\bar{n}$ 
and isoscalar $s\bar{s}$ mesons with natural parity; for the corresponding excited states and the 
other channels we do not have enough bound states with $J=2,3$ to probe for trajectories. 
One immediately notes that, indeed, the sequence $J^{PC}=1^{--}, 2^{++}, 3^{--}$ forms an almost linear
trajectory in the $(M^2,J)$-plane. 
This is interesting, since we are working with a model that is
apparently \emph{not} related to a linear rising potential between light quarks. 
Thus, the 
conventional, naive but intuitive explanation for the formation of Regge-trajectories does not apply
in our framework. Nevertheless, we see an (approximate) $\rho$- and $\phi$-meson Regge trajectory 
for our results. The slope of the trajectory is easily extracted. With
\begin{align}
M^2_X(J) = M^2_X(0) + \beta_X J\;,
\end{align} 
we find
\begin{align}
M^2_\rho(0) &= -0.42 \,\,(-0.05)\,\, \mbox{GeV}^2\;, \nonumber\\
\beta_\rho &= \,\,0.99 \,\,(0.62)\,\, \mbox{GeV}^2\;, \nonumber
\end{align}
and 
\begin{align}
M^2_\phi(0) &= \,\,0.05 \,\,(0.36)\,\, \mbox{GeV}^2\;, \nonumber\\
 \beta_\phi &= \,\,1.12 \,\,(0.78)\,\, \mbox{GeV}^2\;, \nonumber
\end{align}
for $X=\rho$ and $X=\phi$ respectively. The two numbers each correspond to the upper (lower)
end of the $\eta$-band of our results. Compared to recent studies of Regge trajectories
based on the $\rho$-meson, $\beta_\rho = 1.19 \pm 0.10$ GeV$^2$ \cite{Masjuan:2012gc} and 
$\beta_\rho = 1.11 \pm 0.01$ GeV$^2$ \cite{Londergan:2013dza}, our number for the slope at the
upper edge of the $\eta$-band is smaller by only about 
ten percent. Recalling that we need to employ an extrapolation procedure in the complex 
momentum plane to extract the bound state mass of the tensor states with an error margin 
of the order of 5-10 $\%$ the agreement is quite good.

We have also checked for Regge trajectories in channels with unnatural parity and found an 
approximate linear trajectory also for the sequence $J^{PC}=1^{++}, 2^{--}, 3^{++}$ based 
on the $a_0$. Again, for the other channels and the excited states we find not enough bound 
states with $J=2,3$. From the discussion in the previous sections we furthermore expect, 
that the slopes and intercepts in these channels may be further off the experimentally extracted 
values, simply because the rainbow-ladder interaction is not good enough in these channels. 
Indeed for the $a_0$-trajectory we find $M^2_{a_0}(0) = 0.20\,\, \mbox{GeV}^2$ and
$\beta_{a_0} = 0.78$ $\mbox{GeV}^2$ for the upper edge of the $\eta$-band, which do not 
agree too well with {\it e.g.} the values found in Ref.~\cite{Ebert:2009ub}, 
$M^2_{a_0}(0) = -0.658 \pm 0.120\,\, \mbox{GeV}^2$ and
$\beta_{a_0} = 1.014 \pm 0.036 \,\,\mbox{GeV}^2$.

\begin{table*}[!th]
\renewcommand{\arraystretch}{1.2}
\caption{Mass spectrum in MeV for isospin degenerate $n\bar{n}$, isoscalar $s\bar{s}$, and 
$I=1/2$ $n\bar{s}$ bound-states. The rainbow ladder result corresponds to $\eta=1.8\pm0.2$, with 
the superscript ${}^\dag$ (${}^\ddag$) indicating $\eta=2.0$ ($\eta=1.6$) only.}
\label{tab:results}
\begin{small}
\setlength{\tabcolsep}{4.5pt}
\begin{tabular}{c|ccc|ccc||c|ccc}
\hline\noalign{\smallskip}
                            & \multicolumn{3}{c|}{$n\bar{n}   $}                                                         & \multicolumn{3}{c||}{$s\bar{s}$}                                           &          & \multicolumn{3}{c}{$n\bar{s}$}  \\
$J^{PC}$                    & $n=0$                                &   $n=1$                            &          $n=2$ & $n=0$                     &   $n=1$                   &          $n=2$    &  $J^P   $&  $n=0$    &$n=1$  & $n=2$ \\
\noalign{\smallskip}\hline\noalign{\smallskip}                                                                                                                                                                                                   
$0^{-+}$                    & $138.1^{+1.3}_{-0.6}$                &   $1103.0^\dag$                    &  $1770.1^\dag$ &  $696.3^{+2.4}_{-1.7}$    &  $1426.3_{-76.6}$         &                   &  \multirow{2}{*}{$0^-$}       &   \multirow{2}{*}{ $496.6^{+5.3}_{-0.9}$} &  \multirow{2}{*}{$1007.6^{+118.3}_{-\phantom{1}57.0}$ } & \multirow{2}{*}{$1435.9$} \\                                                                                                                                                                        
$0^{--}$                    & $828.8^{+66.9}_{-57.1}$              &                                    &                &  $1133.8^{+68.0}_{-50.8}$ &                           &                   &         &     & & \\
$0^{++}$                    & $643.6^{+17.6}_{-37.6}$              &   $1266.9^\dag$                    &  $1769.1^\dag$ &  $1079.4^{+1.7}_{-7.9}$   &  $1643.6^\dag$            &                   &  \multirow{2}{*}{$0^+$}       &  \multirow{2}{*}{ $874.5^{+10.0}_{-22.2}$ } &  \multirow{2}{*}{$1312.5^{+\phantom{1}90.3}_{-143.8}$} & \multirow{2}{*}{} \\                                                                                                                                                                        
$0^{+-}$                    & $1035.5^{+66.8}_{-38.8}$             &                                    &                &  $1386.7^{+68.8}_{-37.9}$ &                           &                   &         &     & & \\
\noalign{\smallskip}\hline                                                                                                                                                                                                     
$1^{-+}$                    & $1043.9_{-37.0}$                     &                                    &                &  $1347.3^{+73.2}_{-43.7}$ &  $1870.1^{\ddag}$         &                   &  \multirow{2}{*}{$1^-$}       &   \multirow{2}{*}{$ 950.1^{+5.5}_{-1.6}$} &  \multirow{2}{*}{$1241.6^{+43.5}_{-27.9}$ } & \multirow{2}{*}{  } \\
$1^{--}$                    & $757.2^{+1.2}_{-0.6}$                & $1022.6^{+\phantom{1}9.2}_{-29.2}$ & $1331.9^\dag$  &  $1087.8^{+1.8}_{-2.2}$   &  $1413.1^{+38.8}_{-42.1}$ &   $1666.9^\dag$   &         &     & & \\                                                                                                                                                                       
$1^{++}$                    & $969.4^{+15.6}_{-23.9}$              & $1188.1^\dag$                      &                &  $1301.0^{+34.7}_{-28.5}$ &  $1591.9^{+181.2}$        &                   &  \multirow{2}{*}{$1^+$}       & \multirow{2}{*}{$1054.1^{+48.7}_{-44.8}$} &  \multirow{2}{*}{} & \multirow{2}{*}{} \\
$1^{+-}$                    & $852.1^{+13.6}_{-\phantom{1}5.2}$    & $1017.4^{+\phantom{1}0.6}_{-21.4}$ & $1345.2^\dag$  &  $1205.1^{+51.8}_{-46.6}$ &  $1372.0^{+34.4}_{-39.5}$ &   $1831.6^\dag$   &         &     & & \\
\noalign{\smallskip}\hline                                                                                                                                                                                                    
$2^{-+}$                    & $1226.5^{+73.9}_{-80.0}$             &                                    &                &  $1513.5^{+90.5}_{-85.0}$ &                           &                   &  \multirow{2}{*}{$2^-$}       &  \multirow{2}{*}{ $1116.2^{+10.9}_{-17.2}$ } &  \multirow{2}{*}{} & \multirow{2}{*}{} \\
$2^{--}$                    & $1202.6^{+140.0}_{-\phantom{1}94.3}$ &                                    &                &  $1484.7^{+76.0}_{-86.0}$ &                           &                   &         &     & & \\
$2^{++}$                    & $1154.8^{+96.5}_{-69.3}$             &                                    &                &  $1431.4^{+72.4}_{-69.3}$ &                           &                   &  \multirow{2}{*}{$2^+$}       &  \multirow{2}{*}{$1209.4^{+32.3}_{-26.6}$} &  \multirow{2}{*}{} & \multirow{2}{*}{} \\
$2^{+-}$                    &                                      &                                    &                &                           &                           &                   &         &     & & \\
\noalign{\smallskip}\hline                                                                                                                                                                                                    
$3^{-+}$                    &                                      &                                    &                &  $1842.5_{-46.6}$         &                           &                   &  \multirow{2}{*}{$3^-$}       & \multirow{2}{*}{ $1646.9^\dag$ }  &  \multirow{2}{*}{} & \multirow{2}{*}{} \\
$3^{--}$                    & $1528.3^{+\phantom{1}71.2}_{-184.2}$ &                                    &                &  $1751.7^{+99.2}_{-94.3}$ &                           &                   &         &     & & \\
$3^{++}$                    & $1510.5^{+\phantom{1}81.6}_{-100.3}$ &                                    &                &  $1770.9^{+91.4}_{-96.1}$ &                           &                   &  \multirow{2}{*}{$3^+$}       & \multirow{2}{*}{ $1673.4^\dag$  } &  \multirow{2}{*}{} & \multirow{2}{*}{} \\
$3^{+-}$                    &  									&                                    &                &  $1849.4_{-43.6}$  &                           &                   &         &     & & \\                                                                                                                                                                      
\noalign{\smallskip}\hline
\end{tabular}
\end{small}
\end{table*}

\section{Summary and conclusions}\label{sec:conclusions}
We presented the covariant decomposition of $J\le3$ quark-antiquark bound states, 
following~\cite{Joos:1962qq,Weinberg:1964cn} and \cite{Zemach:1968zz}. Within the 
rainbow-ladder truncations using a well-established effective interaction we calculated 
the spectrum of light unflavored and strange mesons. Comparison with experiment highlights 
the need to explore truncations beyond that of rainbow-ladder; in particular the effects 
of mixing as well as the introduction of a flavor dependent interaction are needed.
In the language of potentials for the spin-spin and spin-orbit forces we find 
sizable deviations in all channels, where the tensor part of the spin-spin interaction is important.
These are in particular the scalar and axialvector channels. On the other hand, the results
are quantitatively reliable on the five percent level (at least for the upper end of the 
checked $\eta$-band of the interaction parameter) for channels where only the contact part
of the spin-spin interaction plays a role, {\it i.e.} the hyperfine splitting of the
$S$-states, and channels dominated by the spin-orbit force, {\it i.e.} $J^{PC}= 2^{++}, 3^{--}$.
As a consequence, we find that the ground state Regge trajectory based on the $\rho$-meson
agrees with extractions from experiment on the ten percent level. Since our approach is
{\it not} based on a linear rising potential between light quarks, it is interesting that 
we see approximate Regge trajectories in the first place. This sheds some doubt on the
intuitive but naive interpretation of Regge-behavior as originating from color flux tubes.
The alternative mechanism at work in our framework needs to be explored further. 
Future work will also focus on the accessibility of excited states through a proper
treatment of the quark propagator poles in both the quark DSE and meson BSE, in 
addition to an exploration of the heavy-heavy and heavy-light meson spectrum.

\section*{Acknowledgments}
We thank Gernot Eichmann, Walter Heupel and Helios Sanchis-Alepuz for useful discussions. We are grateful 
to Christian Kellermann for valuable contributions in the early stage of the work. We
thank Gernot Eichmann for comments and a critical reading of the manuscript.
This work was supported by the Helmholtz International Center for FAIR within the LOEWE 
program of the State of Hesse, by the BMBF under contract No. 06GI7121, and 
the Austrian Science Fund (FWF) under project number M1333-N16.

\appendix
\numberwithin{equation}{section}

\vspace*{0cm}
\section{Supplementary Table}
In Table \ref{tab:results} we collect together our results for $n\bar{n}$, $s\bar{s}$ and $n\bar{s}$ states.

\end{document}